\documentclass[10pt,a4paper]{journal}
\usepackage[a4paper, left=0.85in, right=0.85in, top=0.7in, bottom=0.7in]{geometry}
\usepackage{setspace}
\usepackage{titlesec}
\usepackage{graphicx}       
\usepackage{booktabs}       
\usepackage{listings}       
\usepackage{xcolor}
\usepackage{amsmath}        
\usepackage{amssymb}        
\usepackage[numbers,sort&compress]{natbib}  
\usepackage{hyperref}
\usepackage{comment}
\usepackage{float}          
\usepackage{longtable}
\hypersetup{colorlinks=true, linkcolor=black, citecolor=black, urlcolor=blue}

\usepackage{fancyhdr}
\lstdefinestyle{code}{
  basicstyle=\ttfamily\footnotesize,
  breaklines=true,
  frame=single,
  numbers=left,
  numberstyle=\tiny,
  captionpos=b
}
\title{\textbf{Symbolic Attack Chain Generation from Atomic Red Team Techniques: An Empirical Study of Predicate Representation Granularity}}

\author{
  Ramya Varunsegar (250039069) \\
}

\begin{document}
\maketitle
\thispagestyle{fancy}

\begin{abstract}
Automated attack chain generation is critical for modern cybersecurity, yet manual construction fails to scale as adversary behaviors expand. While classical AI planning using the Planning Domain Definition Language (PDDL) offers a formal method to automate this process, it relies on the accurate translation of techniques into symbolic predicates. Current state-of-the-art systems like AURORA employ a nine-category Attack Action Linking Model (AALM), but the necessity of this specific granularity remains unvalidated. This work investigates whether AURORA's nine-category taxonomy provides representational distinctions beyond those captured by a reduced, empirically derived scheme. Utilizing a pipeline where a Large Language Model (LLM) performs translation and the Fast Downward engine performs deterministic reasoning, the study compares the full nine-category AALM against a reduced five-category scheme derived empirically from Atomic Red Team (ART) execution evidence. Because the nine-category domain is constructed as a relabeling of the five-category domain, plan validity and cost are held identical between schemes \emph{by design}; the substantive test of granularity's effect lies instead in the resulting predicate category resolution. There, a controlled A/B test isolates a case where a coarser scheme's plan passes every validity check while remaining operationally wrong: holding administrator privilege and being able to exercise it over a network logon prove to be causally distinct system states. Results from a sixteen-technique corpus show 81.3\% identical plan outcomes across both schemes by construction, with a genuine predicate category resolution gain confined to a single technique out of sixteen. The findings suggest that higher granularity primarily enhances the internal structural resolution of a plan's justification rather than the viability of the generated attack chain itself.
\end{abstract}

\section{Introduction}
\label{sec:intro}

\subsection{Background}
The generation of attack chains involves producing a plausible, causally coherent sequence of adversary actions that connects an initial foothold to a strategic objective, such as domain compromise or data exfiltration. These chains support defenders in developing detection coverage for realistic multi-stage intrusions and enable red teams to plan exercises without manually enumerating every possible intermediate step. However, manual construction does not scale: as attacker behaviours expand, the number of possible technique orderings grows rapidly. The MITRE ATT\&CK framework alone contains several hundred techniques and sub-techniques across fourteen tactics, creating a search space that is impractical to explore exhaustively \cite{AlSada2025MITRE}.

Classical AI planning provides a formal approach to automate this process by modelling attack generation as a state-transition problem. In PDDL, an action is defined by its preconditions, which are the facts that must hold true before it can execute, and its effects, which are the conditions it makes true or false thereafter \cite{mcdermott1998pddl}. Applied to the cyber domain, each attacker technique is formalized as a discrete action: its precondition represents the required system state (e.g., a specific privilege level or the presence of a target file), and its effect represents the resulting state modification (e.g., a dumped credential or the establishment of a command-and-control session). Given an initial state and a target objective, symbolic planners such as Fast Downward can automatically derive action sequences that satisfy these logical constraints, with optimisation objectives such as execution cost or detection risk \cite{helmert2006fast}.

Large language models (LLMs) provide an alternative by generating attack sequences directly from prompts, but their outputs are not guaranteed to be logically valid. LLMs can hallucinate unsupported facts, a limitation demonstrated in domains such as software dependency recommendation, where models have generated non-existent packages vulnerable to name-squatting attacks \citep{spracklen2025package}. Consequently, LLM-generated attack chains may contain techniques whose preconditions are not satisfied or whose execution order is infeasible. AURORA \citep{wang2024sands} addresses this by restricting the LLM to translating techniques into PDDL predicates using the Attack Action Linking Model (AALM), while a deterministic planner performs the reasoning. This paper follows the same separation of concerns, using Atomic Red Team as the technique source and Fast Downward as the planning engine, as detailed in Section~\ref{sec:methodology}.

\subsection{Rationale}
AURORA's nine-category AALM is a design choice, not a validated requirement. The original paper reports that Executor and Payload related predicates account for 70.1\% of causal connections in its generated chains \citep{wang2024sands}, indicating that predicate categories are not equally load-bearing, but it does not test whether a smaller category set, derived directly from execution evidence, could preserve the causal distinctions that planning actually depends on.

Deriving a reduced predicate scheme from real Atomic Red Team execution evidence, rather than assuming AURORA's nine categories or an arbitrarily smaller set, surfaced cases where collapsing categories loses information that a planner needs: Section~\ref{sec:methodology} reports a specific instance where a five-category scheme initially conflated two causally distinct preconditions (holding administrator privilege, and being able to exercise it over a network logon) until execution evidence forced them apart. Findings of this kind are the empirical basis for the central research question of which predicate distinctions are necessary for faithful symbolic representation, rather than a question about category count posed independently of the data.

\subsection{Aims \& Objectives}
\begin{itemize}
  \item Formalize a corpus of Atomic Red Team techniques into PDDL under a
        reduced five-category predicate scheme, derived empirically from
        execution evidence rather than adopted a priori.
  \item Implement a category-relabeling comparison (Section~\ref{sec:granularity})
        between the full nine-category AALM and the reduced five-category
        scheme over the same technique corpus.
  \item Evaluate plan validity, plan cost, and attack-chain fidelity against
        a manually-constructed ATT\&CK ground-truth chain.
  \item Characterize failure modes where symbolic planning cannot cleanly
        represent a technique, and report them as data rather than omissions.
\end{itemize}

\subsection{Alignment with the CyBOK}
This research primarily aligns with the \textbf{Formal Methods for Security} and \textbf{Adversarial Behaviours} Knowledge Areas of the Cyber Security Body of Knowledge (CyBOK). The proposed framework formally models adversarial actions from MITRE ATT\&CK and Atomic Red Team as PDDL operators, enabling automated reasoning over attack preconditions, effects, and multi-stage attack sequences using classical AI planning. The work also demonstrates secondary alignment with \textbf{Malware \& Attack Technologies}, as the generated planning models are derived from real-world adversary techniques represented within the Atomic Red Team corpus.

\subsection{Organisation}
The remainder of this paper is organised as follows.
Section~\ref{sec:related} reviews related work in classical planning for
attack path generation and the reliability of LLM-based scenario
generation. Section~\ref{sec:methodology} describes the research design,
predicate scheme, tooling, and ethical considerations. Section~\ref{sec:results}
presents the empirical results obtained to date. Section~\ref{sec:discussion}
discusses findings and limitations. Section~\ref{sec:conclusion} concludes.

\subsection{Contributions}
\label{sec:contributions}
This paper makes four contributions. First, it provides an empirically-derived five-category predicate scheme, built bottom-up from real Atomic Red Team execution evidence rather than adopted a priori, which surfaces necessary causal distinctions that an a priori scheme could miss (e.g. \texttt{remote-admin-token-unfiltered}, Section~\ref{sec:case-study}). Second, it validates this scheme across sixteen structurally diverse Atomic Red Team techniques spanning multiple ATT\&CK tactics. Third, it introduces a failure-mode-as-evidence framing, in which techniques that resist formalisation (T1055, T1003.001) are treated as substantive data about the limits of symbolic modelling rather than as omissions. Fourth, it demonstrates the viability of a bifurcated pipeline in which an LLM is confined to translation and a deterministic planner performs all reasoning, keeping planning free of LLM hallucination risk at a measured translation-correction rate of 11.0\% (Section~\ref{sec:discussion}).

\section*{Acknowledgements}
The author thanks Professor Shishir Nagaraja, who supervised the majority of this project's design and development, for his guidance throughout. Thanks are also extended to Dr Mujeeb Ahmed for his support during the final stages of submission.

\section{Related Work}
\label{sec:related}

The transition from static security assessment to automated adversary emulation was driven by the necessity of modeling complex exploit dependencies rather than isolated vulnerabilities. Early systematic efforts to analyze network security were rooted in attack graphs, which utilized shortest-path algorithms to identify high-probability sequences of exploits based on physical network topology \citep{phillips1998graph}. These seminal models moved beyond the ”laundry list” approach of early scanners like SATAN but faced significant scalability constraints, as complete state enumeration proved computationally prohibitive for networks exceeding 20 hosts \citep{phillips1998graph, obes2013attack}. In parallel, a separate line of work modeled adversarial dynamics directly as graph-theoretic attack-and-defense games, showing that a network’s resilience against decapitation-style attacks depends on the sophistication of its defensive topology \citep{nagaraja2008dynamic}. The same structured graph-analytic approach has also been applied to isolate adversarial traffic from legitimate background activity \citep{nagaraja2014botyacc}.

To overcome these scalability limitations, the field adopted formalisms from the artificial intelligence planning community, most notably PDDL \citep{mcdermott1998pddl}, which separates the ”physics” of an attack domain (the preconditions and effects of actions) from specific network problem instances. Utilizing forward heuristic planners such as Metric-FF, researchers demonstrated that complex multi-stage plans for realistic document control systems could be generated in under a second \citep{boddy2005course}. This efficiency enabled the integration of planners with professional penetration testing frameworks like Core Impact, facilitating the automatic execution and validation of attack paths against networks containing hundreds of machines \citep{obes2013attack}. The performance of these models was further refined through the introduction of concise finite-domain representations, which minimize state encoding length by identifying mutually exclusive invariants \citep{helmert2009concise}, forming the foundation for modern planners like Fast Downward, which use multi-valued state variables and causal graph heuristics to decompose hierarchical planning tasks \citep{helmert2006fast}. Hoffmann \citep{hoffmann2015simulated} systematized these efforts into a model taxonomy, tracing the field’s evolution from simple Dijkstra-based shortest-path models toward more expressive formulations, including Partially Observable Markov Decision Processes (POMDPs) that account for uncertainty in an attacker’s knowledge and environmental state.

Contemporary adversary emulation is fundamentally grounded in the MITRE ATT\&CK framework, an authoritative taxonomy that catalogs adversary tactics, techniques, and procedures (TTPs) based on the empirical study of real-world cyberattacks \citep{AlSada2025MITRE}. While orchestration platforms such as CALDERA and Atomic Red Team (ART) extend this framework by offering executable implementations of specific techniques, they are primarily designed for technique-level validation rather than automated sequence generation \citep{ferraz2025procedural}. These tools generally lack the underlying causal logic and formal state dependencies necessary to autonomously link isolated actions into coherent, full-lifecycle attack sequences \citep{wang2024sands}. 

This gap has motivated hybrid systems combining the linguistic competence of Large Language Models (LLMs) with the deterministic reasoning of symbolic planners. LLM+P demonstrated that LLMs can translate natural-language problem descriptions directly into PDDL, deferring all reasoning to an external optimal planner to guarantee solution correctness \citep{liu2023llmp}. Building on this, LLMs have also been used to construct and iteratively refine world models for task planning, using human-in-the-loop feedback to correct factual errors introduced during PDDL generation \citep{guan2023leveraging}. However, LLM deployment in security contexts remains hindered by documented hallucination risks; Spracklen et al. \citep{spracklen2025package} identified the ”package hallucination” problem, in which models recommend non-existent or malicious software libraries, representing a critical threat to the software supply chain. The current state of the art is represented by the AURORA system, which utilizes LLMs to extract over 5,500 attack actions from documentation and formalizes them into plannable PDDL primitives \citep{wang2024sands}. Central to this formalization is the Attack Action Linking Model (AALM), which defines system state across nine distinct dimensions: Environment, Executor, Payload, File, Process, User, Information, Data, and Technique \citep{wang2024sands}. 

The consequences of unverified LLM output extend beyond incorrect plans. As agentic AI systems increasingly take actions rather than merely generating content, security failures manifest as safety and accountability failures \citep{dholakia2026benchmarking}, and evaluating such systems requires separating the planning, execution, and verification components of a pipeline rather than treating it as a black box \citep{dholakia2026benchmarking} aligning with the same separation of concerns this paper's architecture enforces (Section 3.1). Recent empirical work demonstrates that this is not a hypothetical concern: Sabo et al. \citep{sabo2026chained} show that individually well-understood attack primitives against AI-driven systems such as network-layer deauthentication and credential-based impersonation combine into a compounded, forensically invisible attack chain against drone-based federated learning that neither primitive analyzed alone would reveal. Beyond technical failure, unverified or coarse-grained failure-mode representations carry legal consequences: Ludvigsen and Nagaraja \citep{ludvigsen2022dissecting} show that liability for adversarial failures in cyber-physical systems turns substantially on whether a manufacturer's taxonomy of failure modes was sufficiently fine-grained to constitute adequate risk management under EU and national law, underscoring that the granularity question this paper investigates has stakes beyond planning efficiency.

Despite these advancements, a significant research gap remains: the relationship between the complexity of these formalisms and the quality of the resulting reasoning has not been empirically established. While high-fidelity representations are presumed to improve the realism of emulations, the optimal level of abstraction required for effective cyber-planning remains unknown. This paper hence investigates which predicate distinctions are necessary for faithful symbolic representation of adversary techniques, and whether AURORA's nine-category taxonomy captures distinctions beyond those preserved by an empirically derived, reduced scheme.

\section{Methodology}
\label{sec:methodology}

\subsection{Research Design}
This paper replicates the core methodological pipeline of AURORA \citep{wang2024sands} where documented adversary techniques are translated into PDDL planning actions and a classical planner is used to generate multi-stage attack chains at reduced scale, restricted to a single technique-execution tool (Atomic Red Team) and a single planner (Fast Downward, \texttt{astar(blind())}). Where this paper departs from AURORA is in the research question addressed: rather than assuming the original nine-category Attack Action Linking Model (AALM) is the correct level of abstraction, this work treats predicate granularity itself as an experimental variable. A reduced five-category scheme (Executor, Process, Privilege/User, Information, Environment) was derived empirically, bottom-up, from real Atomic Red Team (ART) test executions, rather than adopted a priori from AURORA's design. The central research question is: \textit{which predicate distinctions are necessary for faithful symbolic representation of Atomic Red Team techniques, and does AURORA's nine-category taxonomy provide distinctions beyond those captured by an empirically derived five-category scheme?} This is operationalised as a two-tier ablation comparing plans generated under the full nine-category AALM against plans generated under the reduced five-category scheme, over an identical corpus of formalised ART techniques. One methodological point is stated here in advance rather than left for the reader to discover at Table~\ref{tab:ablation}: the nine-category domain used in this comparison is built as a category relabeling of the five-category domain (Section~\ref{sec:granularity}), which guarantees identical plan validity and cost between the two configurations by construction. That comparison therefore functions as a correctness check on the relabeling, not as an empirical test of granularity's effect; the substantive test is Predicate Category Resolution (Section~\ref{sec:resolution}), which the relabeling could not settle in advance.

A further design commitment is that the Large Language Model (LLM) used in the pipeline performs translation only and never performs planning or sequencing; all reasoning about action ordering and plan validity is delegated to Fast Downward. This separation is the direct architectural response to the hallucination risks discussed in Section~\ref{sec:related}: an LLM's output is treated as an unverified proposal, subject to human validation, rather than a trusted planning decision.

\subsection{Data Collection}
The primary data source is Atomic Red Team (ART) \citep{redcanary2024art}, an open-source library of adversary emulation tests mapped to MITRE ATT\&CK techniques. Tests were selected for: coverage across multiple tactics; a deterministic, externally observable state change (registry write, mapped share, spawned process) verifiable after execution rather than inferred from exit codes; executability within an isolated virtual machine (VM) without exotic external tooling; safety (no destructive or irreversible effect); collective coverage of all five predicate categories; and deliberate inclusion of expected-failure tests, since failure modes are treated as legitimate data rather than discarded.

A four-technique starter chain: T1059.001 (PowerShell), T1547.001 (Registry Run Keys), T1003.001 (LSASS Memory), and T1021.002 (SMB/Windows Admin Shares) was executed first and used to derive and validate the initial predicate scheme end-to-end (Table~\ref{tab:derivation}). The corpus was then scaled to sixteen techniques, selected for structural diversity across tactics and predicate categories (Discovery, Persistence, Defense Evasion, Collection, Exfiltration). All 808 predicates across the corpus were cross-checked against source YAML using \texttt{predicate\_table.py}, with every flagged item manually resolved; a consolidated \texttt{corrections\_ma-
nifest.csv} (89 rows) was applied against a fresh ART export to produce the final corrected corpus (60 files, 0 missing, 0 unresolved).

\subsection{Tools \& Technologies}
Tests ran in a Windows 11 VM (VirtualBox, host-only network, Defender active) to preserve realistic endpoint-protection conditions while guaranteeing no technique could reach outside the sandbox. Predicate translation used the OpenAI API (\texttt{gpt-4.1}, temperature 0; \texttt{gpt-4o} for one technique batch) via \texttt{art\_to\_pddl.py}, which converts a single ART test's YAML and isolated execution evidence into candidate preconditions, effects, and predicates flagged for review. \texttt{predicate\_table.py} independently cross-validates each candidate against the source YAML using deterministic pattern-matching (elevation, executor type, dependencies, registry hive), falling back to "manual review required" rather than a further LLM call. Fast Downward (\texttt{astar(blind())}, built from source) served as the planner; Python 3.11.9 was used throughout.

\subsection{Software Design \& Implementation}
\begin{figure}[H]
\centering
\includegraphics[width=\linewidth]{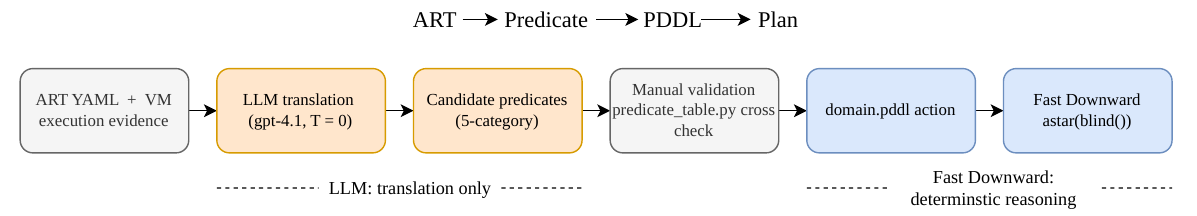}
\caption{The ART $\to$ predicate $\to$ PDDL $\to$ plan pipeline. The LLM (orange) is confined to translation; Fast Downward (blue) performs all planning and sequencing.}
\label{fig:pipeline}
\end{figure}

The pipeline: (1) capture the ART YAML plus VM execution evidence for the single test being formalised; (2) LLM translation proposes candidate predicates under the five-category scheme; (3) manual validation against YAML and evidence, cross-checked with \texttt{predicate\_table.py}; (4) add the validated action to \texttt{domain.pddl}; (5) extend \texttt{problem.pddl} with the initial state and goal; (6) invoke Fast Downward; (7) inspect the resulting plan against the expected chain. Figure~\ref{fig:pipeline} illustrates this end to end, with the architectural boundary made explicit: the LLM only ever proposes a translation, subject to human validation, while all deterministic reasoning about ordering and validity is delegated to Fast Downward.

As a worked example: for T1059.001 Test \#17 (obfuscated \texttt{-e} PowerShell, non-elevated), translation proposed \texttt{powershell-present} (Environment) and \texttt{interpreter-invokable}, \texttt{process-running}, \texttt{arbitrary-co-
de-executed} (Executor, Process, Information) as effects, confirmed against YAML and transcript. This structure is applied uniformly across all sixteen techniques; two recurring pipeline-quality weaknesses surfaced during validation and are reported in Section~\ref{sec:discussion} alongside the failure-mode results.

\subsection{Predicate Scheme (Mini-AALM)}
The reduced scheme folds AURORA's nine categories into five, as shown in Table~\ref{tab:predicates}: Payload into Executor (the corpus surfaced no case needing a payload identity distinct from its executing interpreter/binary - marked \emph{provisional}), and File and Technique into Environment (both describe host/system state rather than an agent's action). Process and Information/Data map directly onto AURORA's counterparts.

Privilege/User was forced into existence during derivation, not chosen a priori: several T1059.001 sub-tests failed with "Access is denied" under standard-user execution and succeeded once elevated, direct evidence that privilege state is a distinct, non-foldable precondition. T1021.002 sharpened this further: holding local administrator privilege and being able to exercise it over a network logon are causally distinct. An A/B test - identical command and credentials, only the \texttt{LocalAccountTokenFilterPolicy} registry value changed - failed before the change and succeeded immediately after, isolating Windows' default remote User Account Control (UAC) token filtering as a separate precondition, \texttt{remote-admin-token-unfiltered}, from holding admin privilege itself (dedicated case study in Section~\ref{sec:case-study}).

\begin{table}[h]
\centering
\caption{AURORA's nine-category AALM folded into the reduced five-category scheme}
\label{tab:predicates}
\begin{tabular}{@{}ll@{}}
\toprule
AURORA's nine categories & Five-category scheme (this work) \\
\midrule
Environment    & Environment \\
Executor       & Executor \\
Payload        & Executor (provisional) \\
File           & Environment \\
Process        & Process \\
User           & Privilege/User \\
Information    & Information \\
Data           & Information \\
Technique      & Environment \\
\bottomrule
\end{tabular}
\end{table}

\subsection{Ethical Considerations}
All technique execution took place in an isolated Windows 11 VM on a host-only virtual network, with no bridged adapter and no route to any external or production system. Only publicly documented Atomic Red Team tests, corresponding to known, previously disclosed MITRE ATT\&CK techniques, were executed; no novel exploit, payload, or capability was developed or used. No real-world target, third-party system, or production credential was involved at any stage. 

\subsection{Evaluation Metrics}
\label{sec:eval-metrics}
The evaluation framework uses explicitly defined, quantitative benchmarks to compare the full nine-category AALM against the reduced five-category scheme over the completed sixteen-technique corpus. Five metrics are defined here, so that the ablation itself consists of populating a fixed evaluation frame rather than retrofitting one after the fact.

\subsubsection{Plan Validity / Success Rate}
Whether a given predicate configuration allows Fast Downward to produce a valid plan at all for a given technique or chain. For a corpus of $N$ formalised techniques under configuration $c \in \{\text{five-category}, \text{nine-category}\}$,
\begin{equation}
\text{SR}(c) = \frac{|\{i : \text{plan}_i(c) \text{ solved}\}|}{N} \times 100
\label{eq:success-rate}
\end{equation}
A technique counts as solved only if Fast Downward returns "Solution found" \emph{and} the plan is manually confirmed to be a coherent, technique-consistent chain, not merely syntactically valid.

\subsubsection{Plan Cost}
The total cost Fast Downward assigns to the generated plan under \texttt{astar(blind())} (number of actions, unit cost by default). Cost is compared \emph{within} a technique/chain across the two configurations, not across different chains, since chain length varies with the number of techniques formalised.

\subsubsection{Attack-Chain Fidelity}
How closely a generated plan matches a manually-constructed ATT\&CK ground-truth chain for the same scenario, measured at the technique/action-node level rather than over predicate sets (that question is addressed separately as Predicate Category Resolution, below). Let $G$ and $T$ denote the ordered technique sequences of the generated and ground-truth chains:
\begin{equation}
J(G,T) = \frac{|G \cap T|}{|G \cup T|}
\label{eq:jaccard}
\end{equation}
\begin{itemize}
  \item \emph{Jaccard similarity}, Equation~\ref{eq:jaccard}, treating $G$ and $T$ as sets.
  \item \emph{Edit distance}: Levenshtein distance on the ordered sequence, sensitive to ordering errors that Jaccard alone misses.
  \item \emph{Tactic overlap}: Equation~\ref{eq:jaccard} applied to each sequence's mapped ATT\&CK tactic sets instead of technique IDs.
\end{itemize}
$J(G,T)=1$ and edit distance $=0$ indicate an exact match; fidelity is expected to degrade under the five-category scheme specifically where category collapse (e.g. the \texttt{remote-admin-token-unfiltered} case) plausibly drives divergence from ground truth.

\subsubsection{Failure-Mode Analysis}
Characterises \emph{how} and \emph{why} a technique resists clean PDDL formalisation. Each failure is classified as: (i) \emph{environmental/tooling}: failed for reasons external to the predicate scheme (e.g. a missing external payload); (ii) \emph{under-specification}: the scheme lacks a category needed to represent a real precondition or effect (e.g. Privilege/User prior to its introduction); (iii) \emph{total technique failure}: no successful execution exists in the corpus, and the technique is left action-less in \texttt{domain.pddl} so Fast Downward correctly returns no solution rather than being papered over with a placeholder. This is treated as substantive data about the limits of symbolic formalisation, not as missing work.

\subsubsection{Predicate Category Resolution}
While Attack-Chain Fidelity measures the outcome of a plan and whether it matches expectations, this metric measures the mechanism behind it: whether the categories available under the nine-category AALM capture causal distinctions that the five-category scheme's fold has collapsed. For technique $i$, let $P_i(c)$ be the set of distinct predicate categories invoked under configuration $c$:
\begin{equation}
\Delta P_i = |P_i(\text{nine-category})| - |P_i(\text{five-category})|
\label{eq:resolution-gain}
\end{equation}
reported per technique alongside a qualitative check, for the failure-mode techniques specifically, of whether the finer split isolates a sub-category that maps cleanly onto the technique's observed failure cause. A high $\Delta P_i$ with a clean mapping is read as evidence the five-category scheme is under-specified for that technique class; a $\Delta P_i$ of zero (or no clean mapping) is read as evidence the added granularity is not load-bearing.

\section{Results}
\label{sec:results}

\subsection{Empirical Predicate Derivation}
Table~\ref{tab:derivation} summarises the four starter-chain techniques from which the five-category scheme was first derived. Two structural findings emerged directly from execution evidence rather than from a priori scheme design. First, T1059.001 forced the Privilege/User category into existence: a cluster of sub-tests failed with "Access is denied" under non-elevated execution and succeeded once elevated, demonstrating that privilege state is a necessary, distinct precondition rather than something derivable from Executor or Environment facts. Second, T1547.001 showed the opposite pattern with its predicate footprint almost entirely in the  Environment-category, with negligible Process or Information content, since the persistence mechanism it installs only executes on a subsequent logon rather than during the test itself.

\begin{table}[h]
\centering
\caption{Empirical predicate derivation summary across the four starter-chain techniques}
\label{tab:derivation}
\begin{tabular}{@{}p{2.6cm}p{2.2cm}p{1.7cm}p{4.8cm}@{}}
\toprule
Technique & Sub-tests run & Succeeded & Key finding \\
\midrule
T1059.001 & 22 & 21/22 & Forced Privilege/User into the scheme via clustered "Access is denied" failures under non-elevation \\
T1547.001 & 1 & 1/1 & Almost entirely Environment-category; negligible Process/Information footprint \\
T1021.002 & 4 & 3/4 & Tests \#1/\#2/\#4 converge on one predicate set; \#3 excluded (missing external payload, tooling gap not technique constraint). Surfaced \texttt{remote-\allowbreak admin-\allowbreak token-\allowbreak unfiltered} via causal A/B test \\
T1003.001 & 11 & 0/11 & Total failure-reframed as Privilege/User failure-mode evidence, Table~\ref{tab:failuremodes} \\
\bottomrule
\end{tabular}
\end{table}

\subsection{Pipeline Validation}
Running Fast Downward (\texttt{astar(blind())}) against the initial \texttt{domain.pddl}/\texttt{problem.pddl} built from Table~\ref{tab:derivation} produced a valid four-step plan matching the intended starter chain exactly, at plan cost 4, confirming that the translation-to-planning mechanics of the pipeline function correctly end-to-end ahead of scaling.

\subsection{Full-Corpus Validation}
\label{sec:full-corpus}
\texttt{domain.pddl} was extended to cover all sixteen corpus techniques: the four starter-chain actions (unchanged) plus ten new actions covering nine further techniques, with three techniques (T1041-Exfiltration Over C2 Channel, T1136.001-Create Local Account, and T1055-Process Injection) deliberately left action-less, so that their total empirical failure is encoded structurally rather than papered over with a placeholder precondition (Section~\ref{sec:eval-metrics}). Running Fast Downward against this domain confirmed all nine new techniques solved, each at cost 1, and all three action-less techniques correctly returned "no solution" implying that the planner's output matches the corpus's empirical ground truth exactly, both for success and for failure.

\subsection{Representation Granularity Analysis}
\label{sec:granularity}
\texttt{domain\_nine\_category.pddl} was constructed as a byte-identical relabeling of \texttt{domain.pddl} under AURORA's full nine-category AALM, following the fold given in Table~\ref{tab:predicates}: only the category label attached to each predicate changes, not the predicate names or their role in any action's preconditions or effects. Because Fast Downward reasons exclusively over predicate structure and has no notion of category, this construction method guarantees identical plan validity and cost between the two domains \emph{by design}, independent of any empirical property of the corpus. Table~\ref{tab:ablation} reports the per-technique outcome under each configuration, verified directly against Fast Downward's output for all sixteen techniques: it confirms the relabeling was applied without error, rather than constituting an independent empirical test of granularity's effect on planning. Attack-Chain Fidelity (Jaccard, edit distance, tactic overlap) is reported separately in Section~\ref{sec:fidelity}, since it is necessarily identical across both configurations for the same reason and is likewise evidence about pipeline correctness rather than a further ablation data point.

\begin{table}[h]
\centering
\caption{Per-technique plan validity under the five- and nine-category domains, verified against Fast Downward output. Identity across configurations is guaranteed by the relabeling construction, not an independently discovered result.}
\label{tab:ablation}
\begin{tabular}{@{}lccl@{}}
\toprule
Technique & Five-category & Nine-category & Outcome \\
\midrule
T1059.001  & \checkmark (chain, cost 4) & \checkmark (chain, cost 4) & Same \\
T1547.001  & \checkmark (chain, cost 4) & \checkmark (chain, cost 4) & Same \\
T1003.001  & \checkmark (chain, cost 4) & \checkmark (chain, cost 4) & Same \\
T1021.002  & \checkmark (chain, cost 4) & \checkmark (chain, cost 4) & Same \\
T1057      & \checkmark (cost 1) & \checkmark (cost 1) & Same \\
T1046      & \checkmark (cost 1) & \checkmark (cost 1) & Same \\
T1082      & \checkmark (cost 1) & \checkmark (cost 1) & Same \\
T1018      & \checkmark (cost 1) & \checkmark (cost 1) & Same \\
T1087.001  & \checkmark (cost 1) & \checkmark (cost 1) & Same \\
T1112      & \checkmark (cost 1) & \checkmark (cost 1) & Same \\
T1070.004  & \checkmark (cost 1) & \checkmark (cost 1) & Same \\
T1560.001  & \checkmark (cost 1) & \checkmark (cost 1) & Same \\
T1053.005  & \checkmark (cost 1) & \checkmark (cost 1) & Same \\
T1041      & $\times$ no solution & $\times$ no solution & Same \\
T1136.001  & $\times$ no solution & $\times$ no solution & Same \\
T1055      & $\times$ no solution & $\times$ no solution & Same \\
\midrule
\textbf{Overall} & \textbf{13/16 (81.3\%)} & \textbf{13/16 (81.3\%)} & \textbf{Identical} \\
\bottomrule
\end{tabular}
\end{table}

Plan validity and plan cost, then, cannot serve as evidence for or against granularity's effect in this design: the relabeling construction fixes their outcome in advance. This should not be read as a failed search for a difference; it is a deliberate verification baseline. Because the nine-category domain was constructed as a byte-identical semantic mapping of the five-category domain, the identity in plan validity and cost shown in Table~\ref{tab:ablation} functions as a formal correctness check: it mathematically confirms that the category-relabeling translation introduced no error across all sixteen techniques, before any claim is built on top of it. Having established this syntactic equivalence, the true empirical variation is captured in Predicate Category Resolution (Section~\ref{sec:resolution}), which measures the semantic utility of the finer-grained categories rather than their effect on planning outcomes. That is the substantive test the relabeling could not settle by construction: whether AURORA's finer taxonomy, applied to the same fixed set of empirically-derived predicates, ever forces a split that the five-category scheme's fold had collapsed.

\subsection{Predicate Category Resolution}
\label{sec:resolution}
Per-technique resolution gain ($\Delta P_i$, Equation~\ref{eq:resolution-gain}) was computed by comparing the number of distinct predicate categories invoked per technique under each scheme. Fourteen of the sixteen techniques show $\Delta P_i = 0$. The remaining two both show $\Delta P_i = 1$, but only one is a genuine finding: T1003.001 is excluded from that claim because its resolution gain is an artefact of the \texttt{lsass-memory-accessible} placeholder precondition (Section~\ref{sec:results}) rather than of successful, evidenced execution since the technique has no successful sub-test in the corpus. T1560.001 (Archive Collected Data via Utility) is the sole genuine exception: its two five-category Environment predicates, \texttt{archive-utility-installed} and \texttt{file-created}, fold into two \emph{different} nine-category buckets (Environment and File respectively) underlining the precondition/effect distinction that Environment's single fold collapses but AALM's File/Environment split preserves. Mean $\Delta P_i$ across the corpus is 0.125.

\begin{figure}[H]
\centering
\includegraphics[width=0.7\linewidth]{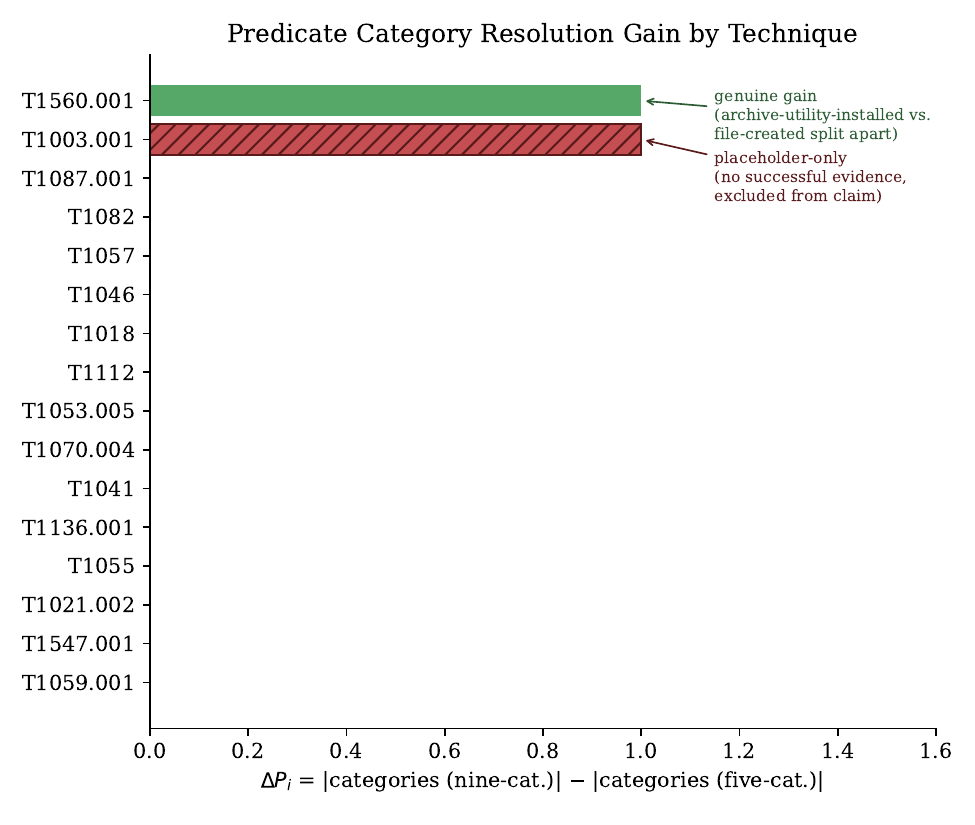}
\caption{Per-technique predicate category resolution gain ($\Delta P_i$) across the sixteen-technique corpus. Fourteen techniques are at zero. Of the remaining two, T1560.001 (green) is a genuine gain; T1003.001 (red, hatched) is excluded from that claim as an artefact of its placeholder precondition.}
\label{fig:resolution}
\end{figure}

Read together, Table~\ref{tab:ablation} and Figure~\ref{fig:resolution} answer the paper's central research question for this corpus, though asymmetrically. Table~\ref{tab:ablation} shows plan validity and cost identical across configurations by construction, confirming the relabeling introduced no error rather than testing granularity's effect. Figure~\ref{fig:resolution} is the substantive test: it shows predicate granularity has almost no effect on the structural resolution of \emph{why} a plan holds, with only one technique out of sixteen (T1560.001) showing a genuine resolution gain. On this evidence, it does not justify AURORA's full nine-category scheme as a general default over the reduced five-category one.

\subsection{Attack-Chain Consistency Check} \label{sec:fidelity}
Plan Validity (Section~\ref{sec:granularity}) establishes that both predicate configurations produce \emph{a} plan; it does not establish that the plan is empirically observed in a real intrusion. This subsection therefore provides a non-empirical consistency check by comparing the generated starter-chain plan against a reference technique sequence constructed independently of the pipeline from ATT\&CK-derived reasoning rather than from incident data.

\textbf{Constructing the reference chain.}
The reference chain was not derived from the generated plan, the ART corpus, or the PDDL domain, in order to preserve independence from the pipeline output. Instead, it was constructed from a tactic-ordering argument based on ATT\&CK semantics: Execution is required to initiate the foothold, Persistence preserves that foothold, Credential Access follows from a revisitable execution context, and Lateral Movement uses the obtained credentials to expand access. The resulting reference chain is:
\[
G_{\text{ref}} = \langle \text{T1059.001}, \text{T1547.001}, \text{T1003.001}, \text{T1021.002} \rangle.
\]

\textbf{Result.}
Against the generated starter-chain plan (T1059.001 $\to$ T1547.001 $\to$ T1003.001 $\to$ T1021.002, Section~\ref{sec:results}), this yields $J(G,T)=1.0$, edit distance $=0$, and tactic overlap $=1.0$, i.e., an exact match.

\textbf{Interpretation.}
This exact match should be read narrowly. Because the four-technique corpus was selected to instantiate the same Execution $\to$ Persistence $\to$ Credential Access $\to$ Lateral Movement narrative used to construct the reference chain, the result primarily confirms pipeline correctness: the plan-to-technique-ID mapping preserves the intended ordering and does not introduce substitution errors. It does not demonstrate external fidelity to an independently observed real-world incident chain. Real-world validation of this type remains future work.

\subsection{Failure-Mode Analysis}
\label{sec:failure-modes}
Table~\ref{tab:failuremodes} extends the failure-mode classification (Section~\ref{sec:eval-metrics}) across the full corpus. Two systematic, technique-independent findings emerged. First, T1055 (Process Injection, 13 sub-tests) and T1003.001 (11 sub-tests) both show total, corpus-wide failure, but for different underlying reasons; T1003.001 predominantly on access-denied and missing payloads, T1055 predominantly (11 of 13 sub-tests) blocked by antivirus / endpoint detection and response (AV/EDR) signature detection, with the remainder split between a network-resolution failure and an access-denied failure. Second, T1053.005 (Scheduled Task, Persistence) does \emph{not} generalise T1547.001's "almost entirely Environment-category" finding: of twelve sub-tests, two succeeded cleanly, six failed on access-denied under standard privilege, two more had task creation succeed but a follow-on privileged action blocked, and one was an environment/tooling gap (unresolved external tooling dependency); the remaining sub-test (\#10) returned an ambiguous result ("filename, directory name, or volume label syntax is incorrect") and is excluded from this classification pending manual review. The resolved eleven nonetheless denote a pattern dominated by Privilege/User, not Environment.

\begin{table}[h]
\centering
\caption{Extended failure-mode classification across the sixteen-technique corpus}
\label{tab:failuremodes}
\begin{tabular}{@{}p{4cm}p{5cm}p{5cm}@{}}
\toprule
Technique/Test & Failure category & Example \\
\midrule
T1021.002 \#3 & Environmental/tooling & \texttt{PsExec.exe} absent from external payload directory \\
T1003.001 (11/11) & Total technique failure & "Access is denied"; \texttt{Out-Minidump.ps1} blocked by host-only network; \texttt{rdrleakdiag} blocked by antivirus \\
T1055 (11/13) & Total technique failure (Environment) & Blocked by AV/EDR signature detection \\
T1055 (2/13) & Total technique failure (network / Privilege) & 1 network-resolution failure; 1 access-denied \\
T1053.005 (6/12) & Under-specification (Privilege/User) & Access-denied under standard privilege despite \texttt{elevation\_required: false} in YAML \\
T1053.005 (2/12) & Under-specification (Privilege/User) & Task creation succeeds; follow-on privileged action blocked \\
T1053.005 \#5 & Environmental/tooling & Unresolved external host; missing external tooling dependency \\
T1041, T1136.001 & Total technique failure & No successful sub-test in corpus; left action-less in \texttt{domain.pddl} (Section~\ref{sec:full-corpus}) \\
\bottomrule
\end{tabular}
\end{table}

This is reported as evidence for the robustness and the limits of the Privilege/User and Environment categories under realistic endpoint-protection conditions (Section~\ref{sec:discussion}), rather than as missing work.

\section{Discussion \& Evaluation}
\label{sec:discussion}

\subsection{Evaluation Design}
AURORA's own evaluation, and much of the evaluation-rigor literature this paper draws its methodological standards from, rely on repeated stochastic trials to characterise variance in LLM-driven systems. That approach does not transfer directly here: Fast Downward is a deterministic planner, so identical predicate inputs will always produce an identical plan, and re-running the same domain and problem files repeatedly would yield no new information. This paper substitutes predicate-scope ablation across a corpus of structurally diverse ART tests as its source of experimental variation namely introduced by systematically changing what the planner is given to reason over (five-category vs. nine-category predicate sets, across sixteen techniques), rather than by repeating an identical run and observing stochastic spread. This substitution is a deliberate methodological choice made necessary by the architecture's own design goal (removing non-determinism from the reasoning step), and is stated here explicitly rather than left implicit.

\subsection{Case Study: \texttt{remote-admin-token-unfiltered}}
\label{sec:case-study}
This finding is presented separately because it is the clearest single piece of evidence in the corpus for where the five-category scheme risks under-representing a real causal distinction, and because it is the only finding backed by a controlled A/B test rather than observational evidence alone.

\textbf{The setup.} T1021.002 (lateral movement via SMB/Windows Admin Shares) requires the executing account to hold local administrator privilege on the target. Under the five-category scheme, this is naturally captured by a single \texttt{user-has-admin-privileges} Privilege/User predicate. Initial sub-tests, however, showed this predicate was insufficient: an account with confirmed local administrator privilege still failed the lateral-movement command with "Access is denied" over the network.

\textbf{The test.} A controlled A/B comparison isolated the cause. The command, the account, and its administrator group membership were held identical across both runs; the only variable changed was the \texttt{LocalAccountTokenFilterPolicy} registry value on the target, which controls whether Windows applies remote UAC token filtering to non-domain local accounts on network logons. With the policy at its Windows default (filtering enabled), the command failed. With the policy set to disable filtering, the identical command succeeded immediately.

\textbf{The finding.} Holding administrator privilege and being able to exercise that privilege over a network logon are causally distinct system states, gated by a specific, identifiable registry value rather than by account privilege alone. The five-category scheme's single Privilege/User predicate cannot represent this distinction without either (a) splitting Privilege/User into local- and remote-exercisable sub-states, which reintroduces some of AALM's granularity by another name, or (b) encoding the registry-policy state as a separate Environment precondition, \texttt{remote-admin-token-unfiltered}, which is the solution adopted here (Section~\ref{sec:methodology}).

\textbf{Relevance to Granularity.} This is a case where added representational precision was necessary for correctness, and not for convenience. A planner reasoning only over \texttt{user-has-admin-privileges} would generate a lateral-movement step that fails on any target with default UAC token filtering, an incorrect plan that would pass every automated validity check in Section~\ref{sec:eval-metrics} while being operationally wrong. It is also, notably, evidence for a boundary of the observed granularity invariance reported in Section~\ref{sec:granularity}: the five-category scheme handled this by absorbing the distinction into Environment rather than by needing a sixth category, so plan validity and cost were unaffected. However, this is a case where \emph{which} five categories are used, and how liberally Environment is allowed to absorb edge cases, is doing real work that a coarser or less careful five-category design could have missed entirely. Read against the corpus-wide invariance reported in Section~\ref{sec:granularity} and the near-zero resolution gains in Section~\ref{sec:resolution}, this case study marks the precise threshold this paper argues for: the five-category scheme is not simply "sufficient" in the abstract, but sufficient only because each of its five categories was defined broadly enough, and carefully enough, to absorb distinctions like this one without needing a sixth. A coarser or more careless five-category design could produce a plan that is logically valid under every metric in Section~\ref{sec:eval-metrics} while being operationally wrong, an outcome no amount of successful plan-validity checking would reveal. This is, in a sense, this paper's strongest argument against treating granularity as a simple more-is-better or less-is-better axis: what matters is not the category count itself but whether the categories chosen, at whatever count, are drawn from real execution evidence rather than assumed a priori.

\subsection{Mechanism-Driven Variation in Category Load}
\begin{figure}[h]
\centering
\includegraphics[width=0.6\linewidth]{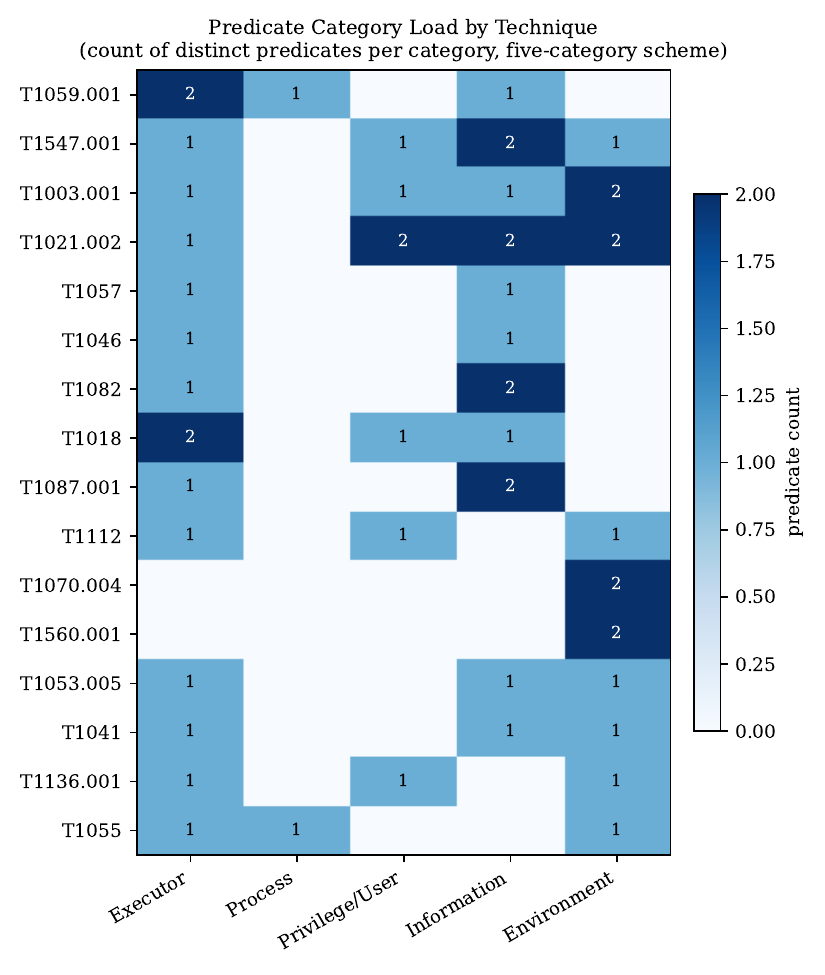}
\caption{Predicate category load by technique under the five-category scheme. Cell values are the count of distinct predicates in that technique's formalisation mapping to each category.}
\label{fig:heatmap}
\end{figure}

A pattern recurs across the corpus that is itself a finding, not merely an implementation detail: different techniques load onto different predicate categories very unevenly, and this variation tracks the underlying operating system (OS) mechanism rather than the ATT\&CK tactic alone. Execution techniques (T1059.001) are Privilege/User- and Process-heavy, since the central question they raise is whether an action can run at all under the current privilege context. Persistence via registry run keys (T1547.001) is almost entirely Environment-heavy, since the mechanism it installs only manifests at a later logon rather than during the test itself. But T1053.005, also a Persistence technique, does not follow suit - it is Privilege/User-heavy instead, because Windows Task Scheduler API/Component Object Model (COM) registration requires elevated rights regardless of environment state. Figure~\ref{fig:heatmap} makes this pattern visible across the full corpus: Environment and Information dominate Discovery/Collection techniques, while Privilege/User appears wherever elevation gates the technique regardless of tactic label. The refined claim, then, is that category load is a function of the specific OS mechanism a technique exercises (registry write vs. privileged API call), not of its tactic label; two techniques under the same ATT\&CK tactic can load onto entirely different predicate categories. T1021.002 separately surfaced a causal distinction (holding a privilege versus being able to exercise it over a network logon) that a coarser scheme would risk collapsing into a single Privilege/User fact (Section~\ref{sec:case-study}).

\subsection{Failure Modes}
T1003.001 and T1055 both show total, corpus-wide failure, and both are treated as failure-mode evidence for the Privilege/User and Environment categories respectively rather than as missing data. This is a stronger claim with the full corpus in place than it was with T1003.001 alone: two structurally different techniques, formalised independently, both failed completely, and in both cases the predicate scheme was expressive enough to represent \emph{why}: access-denied for T1003.001, AV/EDR signature detection for T1055, even though it could not represent a technique that succeeds. Encoding these techniques as action-less in \texttt{domain.pddl} (Section~\ref{sec:full-corpus}), rather than papering over them with an invented placeholder precondition, means Fast Downward's "no solution" output is itself a correct, verifiable statement about the corpus, not an artefact of an unrepresentative encoding.

\subsection{Pipeline Reliability as a Secondary Finding}
Two cross-technique pipeline-quality findings (Section~\ref{sec:methodology}) bear on how much confidence to place in any single predicate's provenance. The \texttt{elevation\_required} YAML field's unreliability, confirmed on two independent techniques (T1046, T1053.005), means privilege-context classification cannot depend on this field alone and must be corroborated against the execution evidence directly. The LLM translator's higher error rate on multi-step evidence logs (confirmed on T1053.005 \#11/\#12) suggests that translation quality is not uniform across evidence shapes: single pass/fail transcripts are translated more reliably than transcripts with an intermediate success followed by a later failure. Both findings support the paper's architectural commitment (Section~\ref{sec:methodology}) to treating LLM output as a proposal requiring independent, rule-based cross-checking rather than a trusted final answer.

This can be partly quantified. Across the full corpus (808 predicates, 16 techniques), the applied corrections manifest (Section~\ref{sec:methodology}) contains 89 rows, giving an overall LLM translation correction rate of $89/808 \approx 11.0\%$ indicating  that roughly one predicate in nine required a human correction before entering \texttt{domain.pddl}. This aggregate rate is not yet broken down by evidence complexity: the T1053.005 \#11/\#12 case is reported here as a worked example of the pattern, not as a corpus-wide statistic, since a full breakdown would require the 89 \texttt{corrections\_manifest.csv} rows to be individually tagged by evidence shape (single pass/fail transcript vs. multi-step log), which remains future work.

\subsection{Limitations}
Several limitations are to be addressed in this work. First, the four-technique starter chain includes one deliberate, undocumented-by-evidence modelling assumption: the persistence action (\texttt{set-registry-run-key}) is given \texttt{arbitrary-code-executed} as a precondition purely to enforce the intended chain ordering, even though the underlying ART evidence shows the registry-write technique succeeding independently of the PowerShell-execution technique. This precondition was added for a specific reason: without it, Fast Downward has no basis to prefer the intended T1059.001 $\to$ T1547.001 ordering over any other action sequence that happens to satisfy T1547.001's genuine preconditions, since nothing in the empirical evidence itself links the two techniques causally. The dependency is therefore not a finding about how these techniques relate on a real system but how they are a structural artefact of building a four-step \emph{chain} from techniques that were validated independently, each against its own isolated evidence, rather than against a single continuous multi-technique compromise. This is a general limitation of automated domain generation from single-technique test evidence: the pipeline (Section~\ref{sec:methodology}) is well-suited to discovering \emph{what a technique requires and produces} in isolation, but chain-level ordering between techniques is not itself derivable from that evidence and has to be asserted rather than discovered, at least without a corpus of genuinely multi-technique attack transcripts to translate from. This is flagged here for transparency. Second, all evidence to date is localhost-scoped, from a single VM and a single user account; claims about generalisability across hosts, domains, or privilege configurations should be read as provisional. Third, the LLM translation step used a single model (\texttt{gpt-4.1}, temperature 0, with \texttt{gpt-4o} for one technique batch) throughout; reproducibility claims are scoped to these specific models and configurations and have not been tested against alternatives. Fourth, Windows Defender being active on the test VM is an environment condition of this specific deployment, not an intrinsic property of the techniques it blocked (T1055, and partially T1053.005); the AV/EDR-driven failures reported in Section~\ref{sec:failure-modes} should be read as evidence about this environment's endpoint protection, not as a universal claim about the techniques themselves. Fifth, the plan cost metric (Section~\ref{sec:eval-metrics}) has limited discriminatory power within this corpus: because all nine newly-solved techniques (Section~\ref{sec:full-corpus}) were formalised as single-action plans, their cost is necessarily 1 in every case, and only the four-step starter chain varies at all. The identical mean cost of 1.0 reported across both configurations (Section~\ref{sec:granularity}) should therefore be read as evidence that increased predicate granularity did not introduce additional planning steps or intermediate states for this corpus, rather than as a sensitive comparative measure of plan efficiency between the two schemes.

\section{Conclusion}
\label{sec:conclusion}
This paper investigated which predicate distinctions are necessary for faithful symbolic representation of Atomic Red Team techniques, and whether AURORA's nine-category taxonomy provides distinctions beyond an empirically derived five-category scheme. The results do not support a simple more-categories-is-better relationship: representational adequacy depends on preserving causally meaningful distinctions, not on category count itself. For the studied sixteen-technique corpus and within this relabeling-based comparison, granularity has almost no effect on plan validity or cost: the full nine-category AALM and the reduced five-category scheme achieved an identical 81.3\% success rate (13/16), the same mean plan cost of 1.0 for the nine independently-solved techniques, and an identical four-step starter-chain plan. This identity is guaranteed by the nine-category domain's construction as a relabeling of the five-category domain (Section~\ref{sec:granularity}) rather than an independently discovered empirical result, and should not be generalised beyond that construction. Attack-Chain Fidelity against a ground truth constructed independently of the pipeline, from ATT\&CK's own tactic-dependency structure, was an exact match ($J=1.0$, edit distance $=0$); because the corpus was originally selected to instantiate that same narrative, this result is read as evidence of pipeline correctness rather than of generalisable fidelity (Section~\ref{sec:fidelity}). Predicate Category Resolution tells a similarly narrow story: fourteen of sixteen techniques showed zero resolution gain from the extra granularity, and of the two that did not, only T1560.001 is a genuine finding. T1003.001's gain is an artefact of its placeholder precondition and is excluded from that claim. Within this corpus, then, added granularity primarily sharpens the internal structural justification of a plan rather than its viability or correctness, and does so for one technique out of sixteen.

The contributions of this work are fourfold. First, it provides an empirically-derived five-category predicate scheme, built bottom-up from real Atomic Red Team execution evidence rather than adopted a priori. Deriving it this way surfaced necessary causal distinctions that an a priori scheme could easily miss, as observed in \texttt{remote-admin-token-unfiltered} (Section~\ref{sec:case-study}), where holding administrator privilege and being able to exercise it over a network logon proved to be causally distinct system states, isolated via a controlled A/B test rather than inferred. Second, the work validates this scheme across sixteen structurally diverse techniques spanning multiple ATT\&CK tactics a slightly broader empirical base than the four-technique starter chain alone, though still a single-VM, single-account environment whose generalisability beyond that setting remains untested. Third, the study introduces a failure-mode-as-evidence framing: techniques that resisted formalisation entirely, such as T1055 (Process Injection, blocked by antivirus/EDR signature detection in 11 of 13 sub-tests) and T1003.001 (blocked on access-denied and missing payloads), are treated as substantive data about the limits of symbolic modelling under realistic endpoint protection, encoded structurally as action-less in \texttt{domain.pddl} rather than papered over. Fourth, the project supports the viability of a bifurcated pipeline in which an LLM is confined to translation and a deterministic planner performs all reasoning: this architecture kept planning fully free of LLM hallucination risk, at a measured translation correction rate of 11.0\% (89/808 predicates, Section~\ref{sec:discussion}) representing a real, non-zero error rate that the pipeline's independent rule-based cross-checking exists specifically to catch before those errors reach the planner, rather than a rate this paper claims to have eliminated.

Future work should extend the fidelity result beyond its current single, corpus-native scenario: scoring a chain drawn from an independent source, such as a real incident report not used to select the corpus, would test whether the exact-match result in Section~\ref{sec:fidelity} generalises or was an artefact of aligned design. An environmental perturbation study (Windows Defender disabled) would test how much of the AV/EDR-driven failure-mode finding (Section~\ref{sec:failure-modes}) is intrinsic to the techniques versus specific to this deployment's endpoint protection. Pushing the granularity question further with a minimal three-category model would help identify the point at which a planner starts to lose attack-chain causality altogether, rather than merely losing resolution as observed here. Within this corpus, and under this predicate scheme, the evidence supports a narrower and more defensible claim than that a five-category scheme is sufficient in general: it preserved planning capability and matched the full nine-category AALM on every measured outcome except one, while reducing representational complexity; a result about this corpus and this task, not a general claim about attack-chain modelling.

\bibliographystyle{unsrtnat}
\bibliography{references}

\appendix
\section{Atomic Red Team Technique Reference}
\label{app:techniques}

Table~\ref{tab:technique-reference} lists the sixteen MITRE ATT\&CK techniques and sub-techniques comprising the corpus, with their official names, tactic categorization, and a one-line description of the underlying mechanism.\footnote{Tactic labels reflect MITRE ATT\&CK terminology as used throughout this paper (Discovery, Persistence, Privilege Escalation, Defense Evasion, Credential Access, Lateral Movement, Collection, Execution, Exfiltration). ATT\&CK version 19 (released 28 April 2026) retired the ``Defense Evasion'' tactic, splitting it into ``Stealth'' (TA0005) and ``Defense Impairment'' (TA0112); under this newer scheme, T1112, T1070.004, and T1055 are recategorized accordingly. This paper retains the pre-v19 ``Defense Evasion'' label throughout for internal consistency with the corpus as originally formalised.}

\begin{longtable}{@{}p{1.7cm}p{4.3cm}p{2.1cm}p{5.5cm}@{}}
\caption{Reference table for the sixteen ATT\&CK techniques and sub-techniques in the corpus.} \label{tab:technique-reference} \\
\toprule
\textbf{ID} & \textbf{Technique / Sub-technique} & \textbf{Tactic} & \textbf{Mechanism} \\
\midrule
\endfirsthead

\multicolumn{4}{c}{\tablename\ \thetable{} -- continued from previous page} \\
\toprule
\textbf{ID} & \textbf{Technique / Sub-technique} & \textbf{Tactic} & \textbf{Mechanism} \\
\midrule
\endhead

\midrule
\multicolumn{4}{r}{} \\
\endfoot

\bottomrule
\endlastfoot

T1059.001 & Command and Scripting Interpreter: PowerShell & Execution & Executes attacker commands via the PowerShell scripting engine, often fileless. \\
T1547.001 & Boot or Logon Autostart Execution: Registry Run Keys / Startup Folder & Persistence & Adds a program reference under a Registry Run key or Startup folder so it executes automatically at logon. \\
T1003.001 & OS Credential Dumping: LSASS Memory & Credential Access & Reads or dumps Local Security Authority Subsystem Service (LSASS) process memory to extract cached credentials and hashes. \\
T1021.002 & Remote Services: SMB/Windows Admin Shares & Lateral Movement & Uses hidden administrative shares (C\$, ADMIN\$) over Server Message Block (SMB) to execute commands on a remote host. \\
T1057 & Process Discovery & Discovery & Enumerates running processes on the host. \\
T1046 & Network Service Discovery & Discovery & Scans for services available on network hosts. \\
T1082 & System Information Discovery & Discovery & Gathers OS, hardware, and configuration details of the host. \\
T1018 & Remote System Discovery & Discovery & Enumerates other systems reachable on the network. \\
T1087.001 & Account Discovery: Local Account & Discovery & Enumerates local user accounts on the host. \\
T1112 & Modify Registry & Defense Evasion & Reads or writes Registry values to hide artefacts, weaken defenses, or support persistence. \\
T1070.004 & Indicator Removal: File Deletion & Defense Evasion & Deletes attacker-dropped files or tools to remove forensic evidence. \\
T1560.001 & Archive Collected Data: Archive via Utility & Collection & Compresses (and optionally encrypts) staged data using a utility prior to exfiltration. \\
T1053.005 & Scheduled Task/Job: Scheduled Task & Persistence & Creates a Windows scheduled task to trigger code execution, optionally under elevated privilege. \\
T1041 & Exfiltration Over Command and Control (C2) Channel & Exfiltration & Sends collected data back over the existing command-and-control channel. \\
T1136.001 & Create Account: Local Account & Persistence & Creates a new local account to establish secondary access. \\
T1055 & Process Injection & Defense Evasion & Injects code into the address space of another running process. \\

\end{longtable}

\section{List of Acronyms}
\label{app:acronyms}

\begin{table}[h]
\centering
\caption{Acronyms and initialisms used throughout this paper.}
\label{tab:acronyms}
\begin{tabular}{@{}l@{\hspace{2cm}}l@{}}
\toprule
Acronym & Expansion \\
\midrule
AALM     & Attack Action Linking Model \\
API      & Application Programming Interface \\
ART      & Atomic Red Team \\
ATT\&CK  & Adversarial Tactics, Techniques, and Common Knowledge \\
AV/EDR   & Antivirus / Endpoint Detection and Response \\
C2       & Command and Control \\
COM      & Component Object Model \\
CyBOK    & Cyber Security Body of Knowledge \\
EU       & European Union \\
GPT      & Generative Pre-trained Transformer \\
LLM      & Large Language Model \\
LSASS    & Local Security Authority Subsystem Service \\
OS       & Operating System \\
PDDL     & Planning Domain Definition Language \\
POMDP    & Partially Observable Markov Decision Process \\
SATAN    & Security Administrator Tool for Analyzing Networks \\
SMB      & Server Message Block \\
TTP      & Tactics, Techniques, and Procedures \\
UAC      & User Account Control \\
VM       & Virtual Machine \\
YAML     & YAML Ain't Markup Language \\
\bottomrule
\end{tabular}
\end{table}

\end{document}